\documentclass [aps,floatfix,subeqn,twocolumn,showpacs,amsmath]{revtex4}
\usepackage{amssymb}
\usepackage{epsfig}
\usepackage{graphicx}
\usepackage{natbib}
\usepackage{picinpar}

\begin{document}

\draft

\title{Pinning and depinning of a classic quasi-one-dimensional
Wigner crystal in the presence of a constriction.}

\author{G. Piacente \cite{giovanni} and F.~M. Peeters \cite{francois}}

\address{Department of Physics, University of Antwerp (Campus Middleheim), Groenenborgerlaan 171,
 B-2020 Antwerpen, Belgium}

\date{\today}

\begin{abstract}

We studied the dynamics of a quasi-one-dimensional chain-like
system of charged particles at low temperature, interacting
through a screened Coulomb potential in the presence of a local
constriction. The response of the system when an external electric
field is applied was investigated. We performed Langevin molecular
dynamics simulations for different values of the driving force and
for different temperatures. We found that the friction together
with the constriction pins the particles up to a critical value of
the driving force. The system can depin \emph{elastically} or
\emph{quasi-elastically} depending on the strength of the
constriction. The elastic (quasi-elastic) depinning is
characterized by a critical exponent $\beta\sim0.66$
($\beta\sim0.95$). The dc conductivity is zero in the pinned
regime, it has non-ohmic characteristics after the activation of
the motion and then it is constant. Furthermore, the dependence of
the conductivity with temperature and strength of the constriction
was investigated in detail. We found interesting differences
between the single and the multi-chain regimes as the temperature
is increased.
\end{abstract}


\maketitle

\section{INTRODUCTION}

In the last years the interest in mesoscopic systems consisting of
interacting particles in low dimensions or confined geometries has
seen a sustained growth. A class of quantum anisotropic systems
exhibiting ``stripe" behavior appears in the quantum Hall regime
\cite{qhestripes}, in charge density waves (CDW) \cite{fogle}, in
manganite oxides and high-T$_c$ superconductors \cite{hightc}
where strong electronic correlations are responsible for the
formation of these inhomogeneous phases. Another class of confined
quasi one-dimensional (Q1D) geometries appears in different fields
of research and some typical and important examples from the
experimental point of view are: ordered electrons on microchannels
filled by liquid Helium \cite{glasson}, microfluidic devices
\cite{whitesides}, colloidal suspensions \cite{zahn} and confined
dusty plasma \cite{chu}. On the atomic scale a chain-like system
was found in compounds such as ${Hg}_{3-\delta}AsF_6$ \cite{brown}
and in low dimensional systems formed on surfaces \cite{segovia}.
These kinds of interacting systems, which tend to form periodic or
ordered structures when the density of particles and the
temperature are low enough (i.e. Wigner crystallization
\cite{wigner,andrei}), can exhibit a remarkable variety of complex
phenomena when they are driven by an external force. Many of these
phenomena, which arise from the interplay between periodicity,
disorder, nonlinearities and driving, are still poorly understood
or even unexplored. For numerous such experimental systems,
transport experiments \cite{charalambous,bhattacharya,ryu} are a
useful way to probe the physics (and sometimes the only way when
direct methods, e.g. imaging, are not available). It is thus an
interesting and challenging problem to obtain a quantitative
description of their non-linear dynamics. One striking property
exhibited by all these systems is pinning, i.e. at low temperature
there is no macroscopic motion unless the applied force $f$
reaches a threshold critical value $f_c$. There is a quite
extensive literature about the dynamical properties near the
depinning threshold \cite{dep1,dep2,dep3}, mostly in the context
of CDW \cite{cdw1,cdw2,cdw3}.

The aim of this paper is to provide a description of the
properties of a Q1D Wigner crystal in the presence of a local
constriction potential, thermal noise and an external driving
force. Most of the previous theoretical and experimental works are
on moving 2D or 3D lattices and glasses (see Ref. \cite{giamarchi}
and references therein). The effect of confinement into a
mesoscopic channel has not yet been deeply investigated.

Our classical model is very ductile because of its scalability and
it is suitable for the description of diverse confined systems, as
electrons on liquid helium, colloids and complex plasmas. We
should stress that the classical approach, which is naturally
valid in the case of colloids and complex plasmas because of the
microscopic size of the particles, is still valid for pure quantum
objects as electrons when they exhibit Wigner crystallization. In
the crystal phase the electrons become localized, and thus
distinguishable. In this case the De Broglie thermal length
$\lambda_D$ is much smaller than the interparticle spacing, hence
the quantum aspect of the original fermionic problem does not play
a crucial role and the classical treatment for the system is an
accurate one. Several generic aspects of the present model without
a constriction and in the linear regime were recently investigated
by the present authors \cite{piacente}.

A narrow channel with a constriction can be readily realized in a
colloidal system or in a dusty plasma. Additionally the problem we
deal with could also be of interest in nanoscale wires.
Flux-line-lattice flow has been studied in a novel superconducting
devices containing straight, nanometer-scale, weak-pinning
channels in a strong-pinning environment \cite{pruy}. By
introducing a constriction into the weak-pinning channels many
features of the model we propose can be investigated
experimentally.

The paper is organized as follows: we first give in Sec. II an
overview of the model and of the numerical methods used. In Sec.
III we describe the zero temperature phase diagram in the absence
of any external driving force, stressing the differences in the
ground state configurations near the constriction. Sec. IV is
devoted to the study of the dynamics of the system, in particular
we concentrate on the velocity vs applied driving force curves and
on the conductivity. In Sec. V we discuss the interplay between
driving force, thermal disorder and constriction potential,
focusing on the difference between the single chain and
multi-chain regime. In Sec. VI we comment on the analogies and
differences with the case of moving lattices and glasses and other
models in which quenched disorder or pinning potentials are
present. Finally we conclude in Sec. VII.

\section{MODEL AND  METHODS}

The system consists of an infinite number of classical identical
particles with charge $q$ and mass $m$, moving in a plane with
coordinates $\vec{r}=(x,y)$. The particles interact through a
screened Coulomb (Yukawa-type) potential, where the screening
length $\lambda$ is an external parameter. We impose a parabolic
confining potential in one direction, namely in the $y$-direction,
and a constriction potential with Lorentzian shape centered in the
origin of the axes. The Hamiltonian of the system is given by:
\begin{equation}
\begin{split}
H=&\frac{q^2}{\epsilon} \sum_{i > j}
\frac{\exp(-|\vec{r_i}-\vec{r_j}|/\lambda)}{|\vec{r_i}-\vec{r_j}|}
\\& +\sum_i \frac{1}{2} m {\omega_0}^2 {y_i}^2 + \sum_i
\frac{V_0}{1+\alpha^2 {x_i}^2},
\end{split}
\end{equation}
\noindent where $\epsilon$ is the dielectric constant of the
medium the particles are moving in, $\omega_0$ measures the
strength of the confining potential, $V_0$ is the maximum of the
potential of the constriction which has a full width at half
maximum $2/\alpha$. Introducing a suitable system of units, the
Hamiltonian can be rewritten in a dimensionless form. We define:
$r_{0} = (2q^2/m\varepsilon \omega_{0}^{2})^{1/3}$ and $E_{0} =
(m\omega_{0}^{2}q^{4}/2\varepsilon ^{2})^{1/3}$ as unity of length
and energy, respectively. After using dimensionless units the
Hamiltonian takes the form:
\begin{equation}
\begin{split}
H'=&\sum_{i > j} \frac{\exp(-\kappa |\vec{r'_i}-\vec{r'_j}|)}
{|\vec{r'_i}-\vec{r'_j}|}\\ &+\sum_i {y'_i}^2+ \sum_i
\frac{V'_0}{1+\alpha'^2 x'^2},
\end{split}
\end{equation}
\noindent where $H^{\prime }=H/E_{0}$, $\kappa =r_{0}/\lambda $,
$\overrightarrow{r}^{\prime }=\overrightarrow{r}/r_{0}$,
$V'_0=V_0/E_0$ and $\alpha'=r_0 \alpha $.

This transformation is particularly interesting because now the
Hamiltonian no longer depends on the mass of the particles, the
dielectric constant of the medium and the frequency of the
parabolic confinement, that is it is independent of the specifics
of the system under  investigation. The dimensionless time and
temperature, which are essential quantities in what follows, are
respectively $T^{\prime }=T/[k_B(m\omega
_{0}^{2}q^{4}/2\varepsilon^{2})^{1/3}]$ and $\tau = t\omega_0$.

The zero temperature configurations for different densities,
namely different number of particles, and different values of the
parameters in the Hamiltonian were calculated by the Monte Carlo
(MC) technique using the standard Metropolis algorithm as it was
done in Ref. \cite{piacente}. We have allowed the system to
approach its equilibrium state at some temperature $T$, after
executing $10^{5}\div 10^{6}$ Monte Carlo steps. In order to reach
the $T=0$ equilibrium configuration the technique of simulated
annealing was used: first the system was heated up and then cooled
down to a very low temperature. We introduced periodic boundary
conditions along the $x$-direction in order to simulate an
infinite long wire. Typically a simulation cell of length $L=100$
(in dimensionless units) centered around the origin of the axis
was used. This choice was motivated by the fact that for larger
$L$ the changes in static and dynamical properties are negligible,
especially for $\alpha'\geq 0.5$ and $\kappa\geq1$.

Because of the presence of the constriction we found many more
metastable states, which complicates the numerical approach. We
will elaborate on this point in the next section.

After reaching the $T=0$ equilibrium configuration, we introduced
an external electrical field in the $x$-direction, or in other
words we considered the effect of an external driving force $f$
and calculate the transport properties of the system. We also
considered the effect of temperature and thermal noise, coupling
the system to additional degrees of freedom \cite{nose,hoover} or
to a heat bath. The Langevin dynamics \cite{kubo} is the most
appropriate one to include such effects. The Langevin equations
for the $x$ and $y$ components of motion in dimensionless units
are respectively:
\begin{subequations}
\begin{equation}
\begin{split}
\frac{d^2{x'_i}}{d{\tau}^2}=&-\gamma\frac{dx'_i}{d\tau}-\frac{1}{2}\sum_j
\frac{\partial}{\partial x'_i}\frac{\exp(-\kappa
|\vec{r'_i}-\vec{r'_j}|)}{|\vec{r'_i}-\vec{r'_j}|}\\&+
 \frac{V'_0 \alpha'^2 x'_i}{(1+ \alpha'^2 {x'_i}^2)^2} + \xi_x(T')
+ f
\end{split}
\end{equation}
\begin{equation}
\begin{split}
\frac{d^2{y'_i}}{d{\tau}^2}=&-\gamma\frac{dy'_i}{d\tau}-\frac{1}{2}\sum_j
\frac{\partial}{\partial y'_i}\frac{\exp(-\kappa
|\vec{r'_i}-\vec{r'_j}|)}{|\vec{r'_i}-\vec{r'_j}|} \\& -y'_i +
\xi_y(T')
\end{split}
\end{equation}
\end{subequations}
\noindent where $\gamma$ is the friction coefficient, which is an
external parameter as well as $\kappa$, and $\vec{\xi}(T')$ is a
random force, reproducing the thermal noise, with zero average and
standard deviation

\begin{equation}
\nonumber \langle \xi_{ir}(\tau)\xi_{js}(\sigma)\rangle=\gamma
T'\delta_{ij}\delta_{rs}\delta(\tau-\sigma).
\end{equation}

\noindent where $(r,s)\equiv(x,y)$. The driving force $f$ and the
random force $\xi$ are measured in units of $m\omega_0^2r_0$,
while the friction coefficient $\gamma$ is measured in units of
$\omega_0$. We used the same simulation cell and the same boundary
conditions as in the case of the MC simulations. It should be
noticed that in the case of colloids \cite{reichhardt} or vortices
in type II superconductors \cite{cao} the motion is overdamped and
Eqs. (3a) and (3b) are simplified: the hydrodynamic interactions
can be neglected and Eqs. (3a) and (3b) reduce to a system of
coupled first order differential equations.

We considered here the more general problem, including also the
hydrodynamic terms. In order to integrate the equations of motion
we used a quasi-symplectic algorithms of \emph{"leap frog"} type
\cite{mannella1} in the form:

\begin{eqnarray}
\nonumber \tilde{r}_i & = & r_i(t)+\frac{\Delta t}{2}v_i(t), \\
\nonumber v_i(t+\Delta t) & = & c_2\bigg[c_1v_i(t)+\Delta t
\frac{\partial H}{\partial r_i}(\tilde{r}_i)+d_1 \eta_i\bigg],\\
\nonumber r_i(t+\Delta t) & = & \tilde{r}_i + \frac{\Delta
t}{2}v_i(t+\Delta t),
\end{eqnarray}

\noindent where $\Delta t$ is the time step, and $\eta_i$ are
Gaussian variables with standard deviation equal to 1 and average
equal to 0; the constants $c_1$, $c_2$ and $d_1$ are given
respectively by:

\begin{eqnarray}
\nonumber c_1 & = & 1- \frac{\gamma \Delta t}{2}, \\
\nonumber c_2 & = & \frac{1}{1+\gamma \Delta t/2},\\
\nonumber d_1 & = & \sqrt{\gamma T' \Delta t}.
\end{eqnarray}

It was shown that this integration scheme has a good stability and
runs rather fast, furthermore it is well behaved in the limit
$\gamma \rightarrow 0$ \cite{mannella2}.

The driving force was increased from zero by small increments. A
time integration step of $\Delta\tau=0.001$ was used and averages
were evaluated during $2 \times 10^5$ steps after $2 \times 10^6$
steps for equilibration.

\section{Ground state configurations}

The ground state configuration is the result of competitive
effects, that is the electrostatic repulsion, the confining
potential that tries to keep the particles close to the $x$ axis
and the Lorentzian constriction potential that prevents the
particles from settling close to the $y$ axis. In Fig. 1  the
contour plots of the sum of both potentials for two different
values of $V'_0$ are shown. Depending on the values of
(increasing) $V'_0$ and (decreasing) $\alpha'$ the saddle point at
$(x,y)=(0,0) $ becomes more pronounced.

\begin{figure}
\begin{center}
\includegraphics[width=8.5cm]{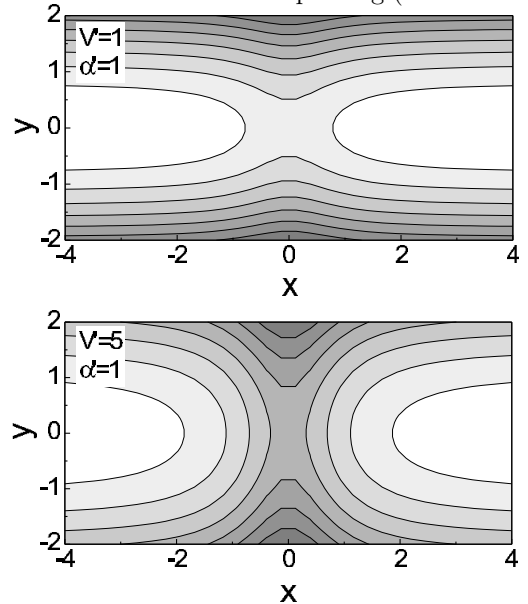}
\caption{Contour plots of the sum of the confinement and
constriction potentials for two different values of the
parameters: (a) $V'_0=1, \alpha'=1$, (b) $V_0'=5, \alpha'=1$.}
\end{center}
\end{figure}

As discussed in Ref. \cite{piacente}, in the absence of any
constriction the charged particles crystallize in a number of
chains. Each chain has the same density resulting in a total
one-dimensional density $\widetilde{n}_{e}$. If $a$ is the
separation between two adjacent particles in the same chain, it is
possible to define a dimensionless linear density
$\widetilde{n}_{e}=lr_{0}/a$, where $l$ is the number of chains.
In the case of multiple chains, in order to have a better packing
(or in other words to minimize the interaction energy by
maximizing the separation among particles in different chains),
the chains are staggered with respect to each other by $a/2$ in
the $x$-direction. For low densities the particles crystallize in
a single chain; with increasing density a ``zig-zag'' (continuous)
transition \cite{zigzag} occurs and the single chain splits into
two chains. Further increasing the density the four-chain
structure is stabilized before the three-chain one. This first
four-chain configuration has a relatively small stability range
after which it transits to a three chain configuration. For higher
values of the density, the four-chain configuration attains again
the lowest energy. Then a further increase of $\widetilde{n}_{e}$
will lead to more chains, that is five, six and so on. The
structural transitions are all discontinuous (i.e. first order),
except the 1 $\rightarrow$ 2 transition.

\begin{widetext}

\begin{figure}
\begin{center}
\includegraphics[width=15cm]{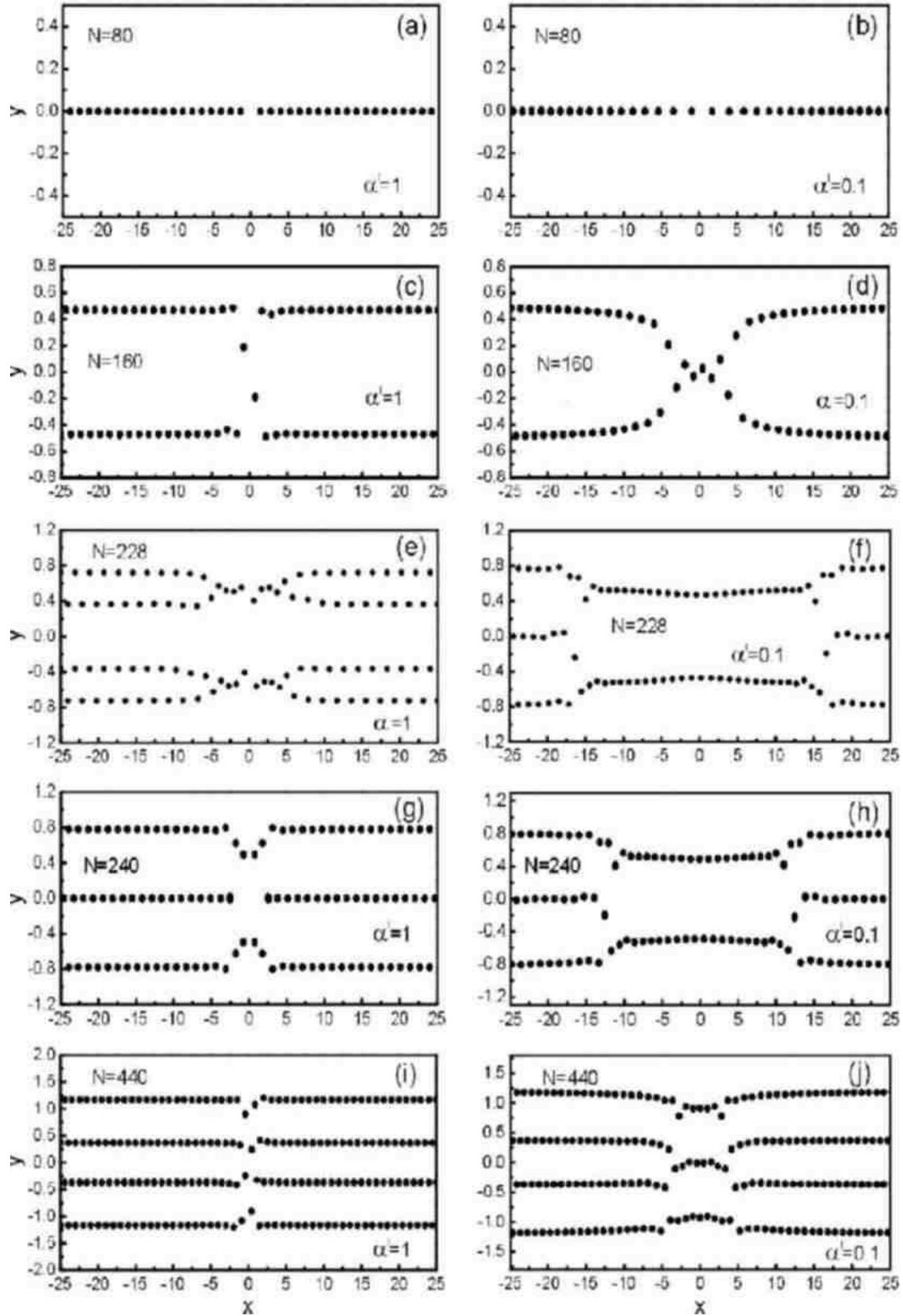}
\caption{The ground state configuration in the center of the
simulation cell for $\kappa=1$, $V'_0=1$ and different values of
the number of particles $N$ in a simulation cell of length $L=100$
and two different values of the constriction width: (a) $N=80,
\alpha'=1$, (b) $N=80, \alpha'=0.1$, (c) $N=160, \alpha'=1$, (d)
$N=160, \alpha'=0.1$, (e) $N=228, \alpha'=1$, (f) $N=228,
\alpha'=0.1$,(g) $N=240, \alpha'=1$, (h) $N=240, \alpha'=0.1$,(i)
$N=440, \alpha'=1$, (j) $N=440, \alpha'=0.1$. In the case of
$\alpha'=1$ the effect of the constriction potential is
significant only around a narrow region around $(x,y)=(0,0)$,
while for $\alpha'=0.1$ it is appreciable all along the lenght of
the simulation cell. It is interesting to notice that for
$\alpha'=1$ the reentrant behavior of Ref. \cite{piacente}, with
the four chain arrangement stabilized before the three chain
arrangement, is still present, while it is absent for
$\alpha'=0.1$. For small values of $\alpha'$ the configurations
are highly inhomogeneous.}
\end{center}
\end{figure}

\end{widetext}

In the presence of the constriction potential the ground state
configurations are modified near the constriction (see Fig. 2),
but the particles are still organized in chains far away from this
constriction. Close to the saddle point the particles do not
arrange themselves in ordered chains. The particles near the
constriction lead to a significant increase of the number of
metastable states. Consequently the procedure of simulated
annealing has to be more accurate than in the case of the absence
of a constriction, which means that several intermediate
temperature steps have to be considered. Sometimes the MC
simulations do not provide us with the ``\emph{exact}" ground
state, as it is seen for instance in Fig. 2 (c), (f) and (j),
where the final configurations are not perfectly symmetric with
respect to $x$ and $y$, while this is expected because of the
symmetry of the Hamiltonian.

When the full width at half maximum of the constriction potential
is short enough ($\alpha' \geq 0.5$), or in other words the effect
of the constriction is significative only in a narrow region
around $x=0$, it is still possible to define a local density
because the system exhibits a homogeneous spacing among charged
particles except in the vicinity of the saddle point. Thus,
excluding these regions, it is still meaningful to consider
$\widetilde{n}_{e}=lr_{0}/a$, where $l$ is the number of chains.
In this case the same chain arrangements, i.e. $1 \rightarrow 2
\rightarrow 4 \rightarrow 3 \rightarrow 4 \rightarrow 5$ and so
on, as in the case where the constriction is absent, is found (see
Fig.2 (a), (c), (e), (g) and (i)) with increasing density, but
with the difference that all the structural transitions are now
discontinuous.

For smaller values of $\alpha$, that is for larger interaction
ranges of the constriction potential, the system is highly
inhomogeneous and even shows coexistence of different chain phases
(see Fig. 2 (d), (f), (h) and (j)). In a certain sense the
constriction introduces a local disorder into the system.

\begin{figure}
\begin{center}
\includegraphics[width=8cm]{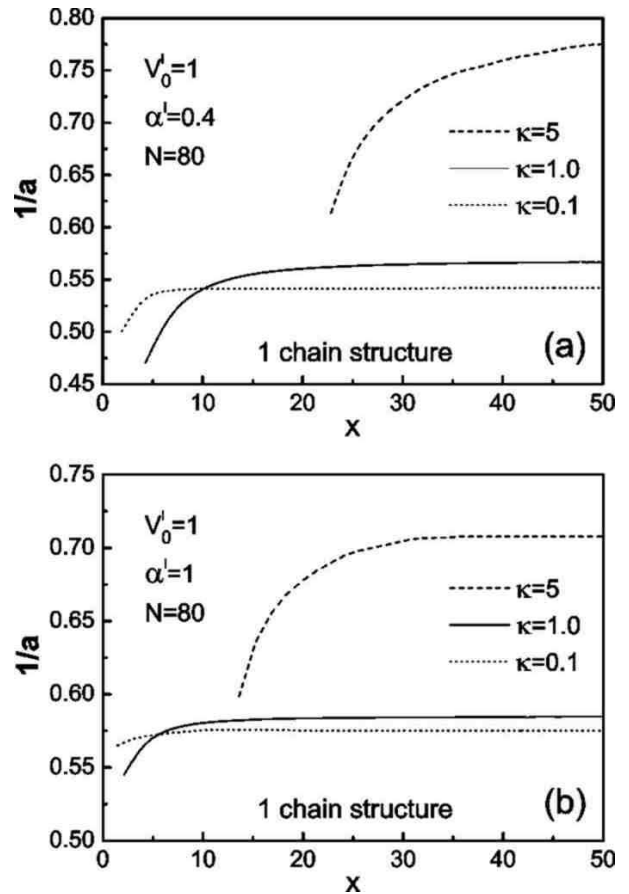}
\caption{The density as a function of the distance from the origin
in the single chain configuration for 80 particles, for a large
(a) and short (b) interaction range of the constriction potential.
Because of the symmetry of the system around $y=0$ only the
results in the right part of the simulation cell are reported.}
\end{center}
\end{figure}

In order to make these affirmations clearer, we plot in Fig. 3 the
inverse interparticle spacing, i.e. the density, in the single
chain configuration as a function of the distance from the origin
of the coordinates for different values of $\kappa$ and $\alpha'$
and $V'_0=1$. It is evident that the density is an increasing
function of the inverse screening length $\kappa$, because the
particles can stay closer together as the electrostatic repulsion
is weaker. For small values of $\alpha'$ the density is an
increasing function of the distance over a large range of
$x$-values (Fig. 2(a)), while for $\alpha'=1$ (Fig. 2(b)) this
range can become very small and the density becomes very quickly
independent of $x$. Notice that for $x\rightarrow\infty$ the chain
density should become independent of the parameters of the
constriction.

\section{Dynamical properties}

When a constant electrical field $E$ is applied to the system in
the $x$-direction, it produces a longitudinal driving force $f$.
The charged particles then are pushed along the direction of the
driving force. In what follows we consider mainly systems for
which $\kappa=1$ and $\gamma=0.2$ , which are typical values for
the inverse screening length and friction respectively,
encountered in complex plasmas \cite {goree}. We also fixed the
value of $\alpha'=1$, that means that we deal with short range
constrictions.

The first obviously important quantity to determine is the
velocity $v'$ as a function of the applied force $f$. In the
absence of thermal fluctuations, i.e. $T=0$, and in the absence of
the constriction potential, Eqs. (3a) and (3b) become:

\begin{subequations}
\begin{equation}
\frac{d{v'^x_i}}{d{\tau}}=-\gamma{v'^x_i}-\frac{1}{2}\sum_j
\frac{\partial}{\partial x'_i}\frac{\exp(-\kappa
|\vec{r'_i}-\vec{r'_j}|)}{|\vec{r'_i}-\vec{r'_j}|}+ f;
\end{equation}
\begin{equation}
\frac{d{v'^y_i}}{d{\tau}}=-\gamma{v'^y}-\frac{1}{2}\sum_j
\frac{\partial}{\partial y'_i}\frac{\exp(-\kappa
|\vec{r'_i}-\vec{r'_j}|)}{|\vec{r'_i}-\vec{r'_j}|} - y'_i.
\end{equation}
\end{subequations}

\noindent Furthermore, because in the equilibrium configuration
the net force acting on every particle, due to electrostatic
repulsion and confinement, is zero, that is

\begin{eqnarray}
\nonumber \frac{1}{2}\sum_j \frac{\partial}{\partial
x'_i}\frac{\exp(-\kappa
|\vec{r_i}-\vec{r_j}|)}{|\vec{r_i}-\vec{r_j}|}=0;\\
\nonumber \frac{1}{2}\sum_j \frac{\partial}{\partial
y'_i}\frac{\exp(-\kappa
|\vec{r_i}-\vec{r_j}|)}{|\vec{r_i}-\vec{r_j}|} + y'_i=0,
\end{eqnarray}
\noindent Eqs. (3a) and (3b) can be ulteriorly simplified and one
obtains the uncoupled equations:

\begin{subequations}
\begin{equation}
\frac{d{v'^x}}{d{\tau}}=-\gamma {v'^x} + f;
\end{equation}
\begin{equation}
\frac{d{v'^y}}{d{\tau}}=-\gamma{v'^y},
\end{equation}
\end{subequations}

\noindent whose stationary solutions are respectively:

\begin{subequations}
\begin{equation}
v'^x=\frac{f}{\gamma};
\end{equation}
\begin{equation}
v'^y=0.
\end{equation}
\end{subequations}

This shows that in the absence of thermal noise and constriction
the total effect of the external driving force is a sliding of the
ordered structure with a drift velocity which is directly
proportional to the driving force and inversely proportional to
the friction. More in general, when the leading term in the
equation of motion is the driving force, one should expect that
the drift velocity is $v'^x=f/\gamma$, or in other words that the
system behaves like a classical two-dimensional Drude conductor
\cite{drude}. This feature has been observed in experiments
\cite{glasson} and in numerical simulations \cite{chen}. In the
presence of a constriction and thermal noise, $v'^x$ is no longer
a linear function of the driving force, as we will discuss in the
next subsections.

\begin{widetext}

\begin{figure}
\begin{center}
\includegraphics[width=17cm]{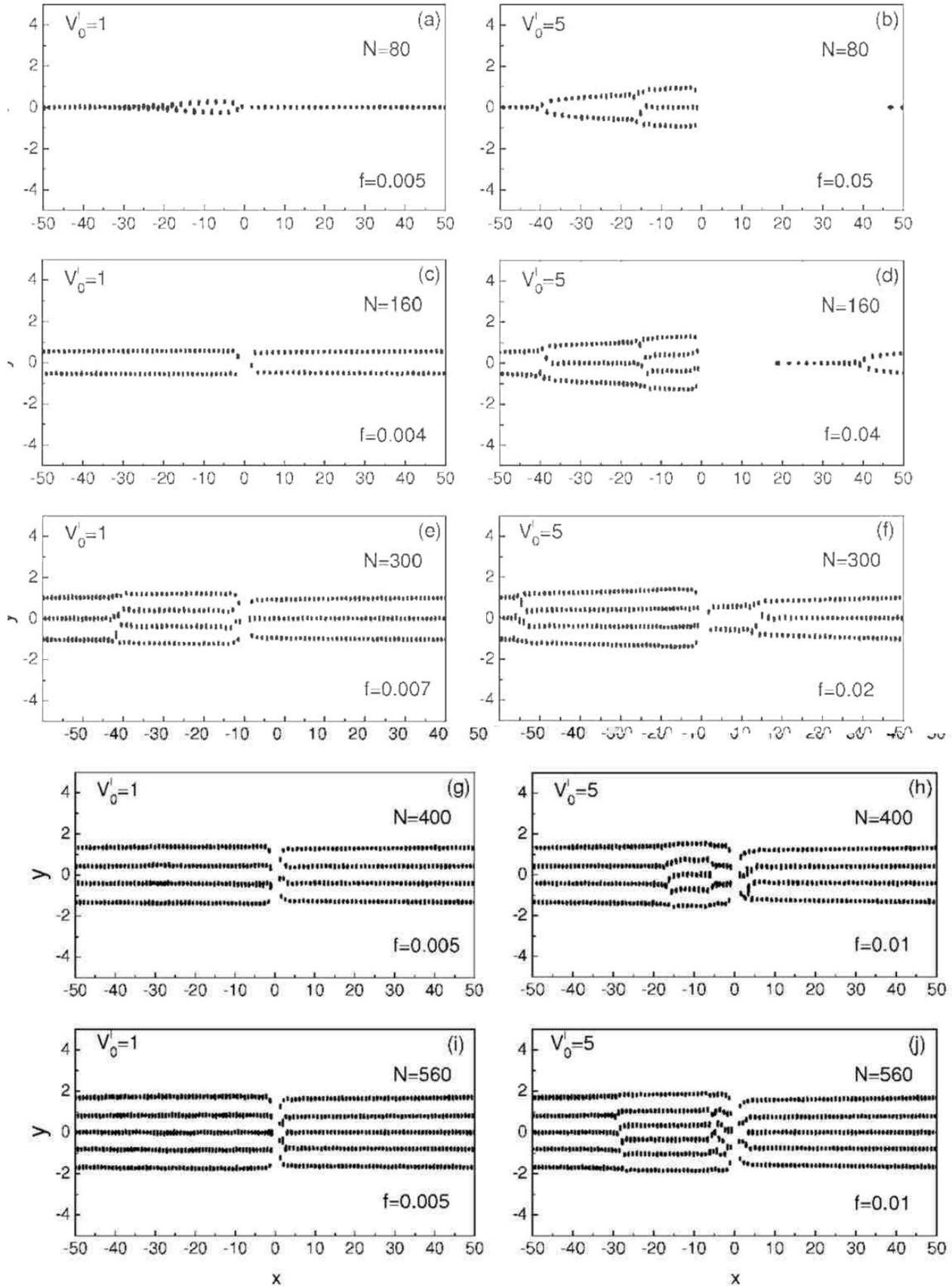}
\caption{Typical trajectories of the particles when the system is
pinned for different number of particles $N$ in a simulation cell
of length $L=100$ and different values of the height of the
constriction barrier. The plots are for a temperature $T'=0.002$,
$\kappa=1$ and $\alpha'=1$.}
\end{center}
\end{figure}

\end{widetext}

\subsection{Pinning}

The system is pinned until the applied driving force reaches a
threshold value $f_c$. The pinned structures in the presence of
the driving force show substantial differences with the ground
state configurations in the absence of a driving force. The
particles move into the direction of the driving force and
accumulate in front of the constriction, that exerts a force,
which is opposite to the driving force. If $f<f_c$, new static
configurations are reached, in which the electrostatic repulsion
and the repulsive force due to the constriction balance the
driving force. The situation is depicted in Fig. 4, where the
driving force is in the positive direction of the $x$ axis. In the
case of low constriction barrier height, the chain structures are
relatively homogeneous, although the inter-chain distance is
smaller at the left than at the right of the constriction because
of the external drive. In the case of high constriction barrier,
the chain structures are no longer homogeneous. As can be seen in
Fig. 4, different number of chains can coexist in the same
configuration. Since the energy to overwhelm the barrier is quite
large, the particles tend mainly to accumulate at $x<0$, which
produces a density gradient and consequently a splitting into a
larger number of chains where the density is larger, because
obviously in such a case the electrostatic repulsion among
particles is larger.

It should be noticed that the nature of pinning for the system
under investigation is different from the pinning often studied in
literature for e.g. colloidal systems and vortex lattices. In
these cases the pinning is the result of some kind of disorder (in
most cases quenched disorder) or, in other words, the effect of
the substrate. It is introduced into the system and modeled as
randomly placed point-like pinning centers producing an attractive
gaussian potential \cite{koshelev,chen,cao1,cao2,brandt} or as
randomly placed parabolic traps \cite{reichhardt}. In our case,
the pinning is the effect of a constraint, the particles have not
enough energy to overcome the constriction barrier and
consequently there is no net motion. We also investigated the case
of negative $V'_0$, i.e. a single Lorentzian potential well. In
that case we still observed pinning, but without the formation of
highly inhomogeneous structures or, in other words, without the
accumulation of many particles in the direction of the driving
force. In the case of a negative Lorentzian potential the same
chain configurations are preserved along the simulation cell
length even in the case of a large depth of the potential well. In
the case of CDW the pinning potential can be either attractive or
repulsive, indeed the the sign of the potential can be converted
by changing the phase of the CDW by $\pi$. In what follows we will
limit ourselves to positive $V'_0$.

In Fig. 4, the trajectories of the particles are reported for a
temperature $T'=0.002$, well below the melting temperature. It is
interesting to study for a fixed number of particles and for a
fixed temperature how by increasing the driving force the
configurations change. The variation of the density along the
constriction is shown in Fig. 5 for different values of the
external driving force for a constriction height of $V'_0=5$ and
width $\alpha'=1$. Increasing the driving force more and more
particles accumulate to the left of the constriction barrier in
the direction of the driving force, corresponding to larger and
larger densities. The density $\widetilde{n}_{e}$ has a
discontinuity at the constriction. For low values of the driving
force, except in the vicinity of the constriction,
$\widetilde{n}_{e}$ is almost constant. But for larger values of
the driving force it is always an increasing function of the
distance along the simulation cell, except close to the
constriction.

\begin{figure}
\begin{center}
\includegraphics[width=8.5cm]{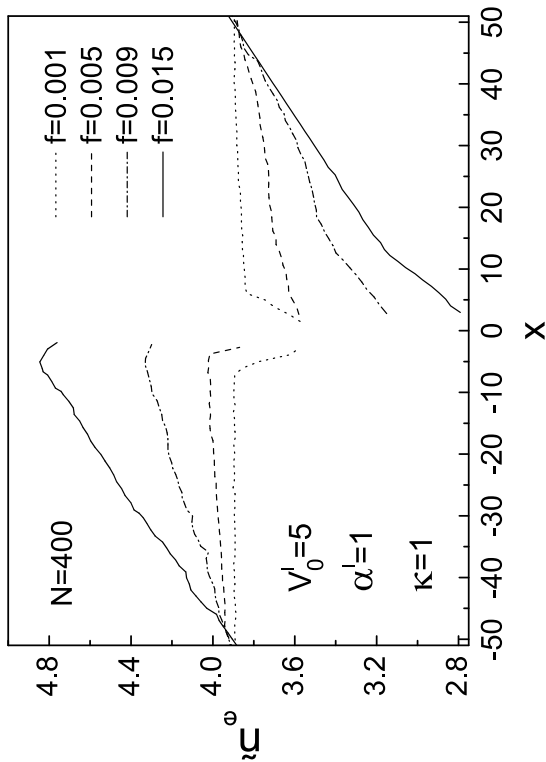}
\caption{The density as a function of the distance along the
simulation cell for different values of the driving force and for
a temperature $T'=0.002$.}
\end{center}
\end{figure}

Increasing $T'$, we observed larger and larger oscillations of the
particles until the system is melted. As already reported in Ref.
\cite{piacente}, also in this case we observed larger oscillations
in the $x$-direction than in the $y$-direction, which is evidently
a combined effect of the confining potential and the nature of the
interparticle interaction. In the case of high temperature, close
to the solid-liquid transition, and mainly in the case of large
$V'_0$ the arrangements of the particles are slightly different
from the ones shown in Fig. 4. Because high values of $V'_0$ in
combination with the driving force produce a density gradient in
the chain-like structures, the melting is not homogeneous, with
the coexistence of solid and liquid regions. It is beyond the aim
of the present paper to discuss how the driving force induces
local melting of the system.

The critical force $f_c$ is evidently a function of the
temperature $T'$, it decreases with increasing temperature, that
is the thermal motion aids the net motion of the particles. The
critical force is also a function of the density, i.e. the number
of particles. In our simulations we observed that for larger
densities $f_c$ becomes smaller.

\subsection{Depinning}

When the driving force $f$ is larger than the threshold $f_c$, the
system exhibits Q1D flow. In Fig. 6 some typical trajectories of
the depinned particles are reported, for $T'=0.002$ and different
values of $f$ just above $f_c$.

\begin{widetext}

\begin{figure}
\begin{center}
\includegraphics[width=16cm]{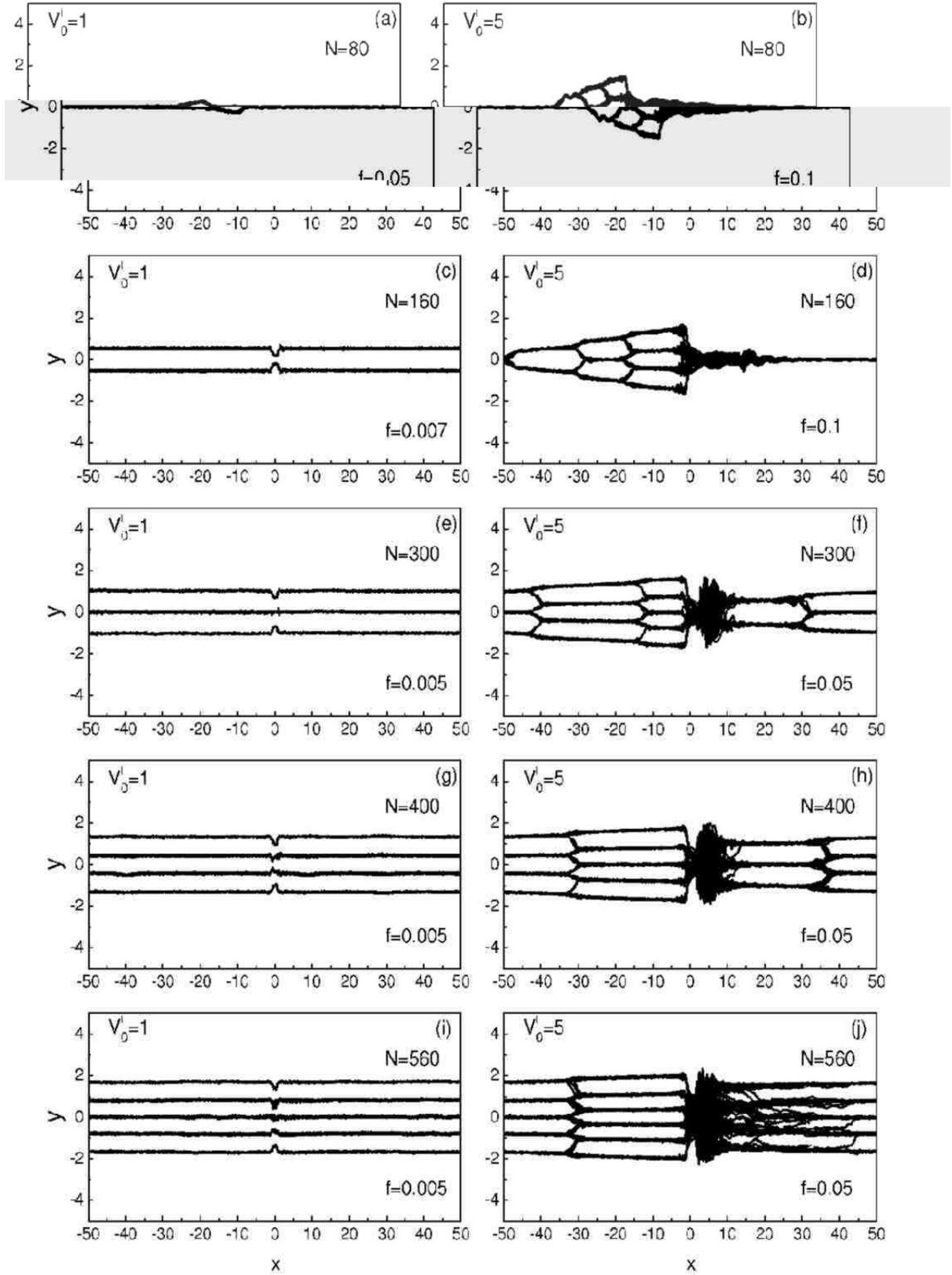}
\caption{Typical trajectories of the particles when the system is
depinned for different densities and different values of the
height of the constriction barrier. The plots are for a
temperature $T'=0.002$, $\kappa=1$ and $\alpha'=1$.}
\end{center}
\end{figure}

\end{widetext}

The first interesting observation is that the driven system does
not break up into pinned and flowing regions, as observed in
experiments and simulations of superconducting vortices
\cite{tanomura,jensen,olson} or colloids \cite{reichhardt}, or, in
other words, the chain-like system under investigation does not
exhibit \emph{plastic depinning}. Once the driving force
overwhelms the critical threshold $f_c$, all the particles move
together. The depinning can be either \emph{elastic} or
\emph{quasi-elastic} depending on the height of the constriction
barrier. In the case of a low barrier ($V'_0=1$ in Fig. 6) the
particles move such that they keep the same neighbors, thus the
system depins elastically. In contrast in the case of a high
barrier ($V'_0=5$ in Fig. 6) a complex net of conducting channels
is activated and the particles move without keeping their
neighbors, that is the depinning is quasi-elastic. The
quasi-elastic depinning is a feature closely related to the low
dimensionality of the system and to the fact that we are
considering a single constriction. We found that quasi-elastic
depinning appears when the strength of the constriction potential
is increased. In other infinite 2D systems of driven particles or
vortices, where several pinning centers are present, a crossover
from elastic to plastic depinning with increasing strength of the
pinning potential occurs (see e.g. Ref. \cite{reichhardt}). We
will focus on this problem in the next subsection.

The region after the constriction barrier, as it is evident from
the trajectory patterns, shows features that deserve a deeper
investigation in the case of high values of $V'_0$. In Fig. 6 for
$V'_0=5$ we found that at the right of the constriction some noise
is present and the particles flow disorderly. In order to explain
this behavior we investigated the distributions of the $x$ and $y$
components of the velocity in narrow strips along the simulation
cell length. We concentrated our attention on a system of $N=400$
particles at $T'=0.002$, with $\kappa=1$, $V'_0=5$ and
$\alpha'=1$. The results are reported in Fig. 7.

It is evident that $v_y$ is always normally distributed with
average equal to zero, as expected, because it receives
contributions mainly from thermal noise, which is gaussian. The
distribution of $v_x$ is still gaussian, but centered around a
value $<v_x>\neq0$ because of the external driving force, except
in the neighborhood of the constriction where the strong
interaction with the barrier gives a non-gaussian profile to it.
What is interesting is the fact that the velocity above the
depinning threshold has a pronounced gradient in the
$x$-direction. From Fig. 7, it is clear that approaching the
constriction from the left side the particles are slowed down;
they receive a sudden acceleration when they pass the constriction
barrier, then the velocity has a maximum in the right neighborhood
of the constriction and finally it slows down again when
approaching the edge of the simulation cell. In order to explain
these highly non linear features, it is helpful to look at the
profile of the force due to the constriction potential (see Fig.
8). This force has a significative magnitude only in a narrow
region around $x=0$. For $x<0$ it acts oppositely to the driving
force, while for $x>0$ it enhances the driving force. There are
two maxima for the intensity of the force located at
$x=\pm1/\sqrt{3}$ and it is zero at the origin of the axis.
Therefore, when the particles approach the constriction they start
to feel this decelerating force and slow down. Because the system
is strongly interacting the deceleration is seen not only in the
left neighborhood of the barrier, but in a wider region. At $x=0$
the force is zero, for $x>0$ close to the constriction the force
quickly increases and adds to the driving force, so the particles
experience a sudden acceleration which produces a large velocity.
After that the particles are accelerated only by the driving force
and start to feel the effect of the particles on the opposite side
of the simulation cell because of the periodic boundary
conditions, so the velocity decreases again. In the case of low
constriction barriers or large driving forces the $x$ components
of the velocity are more homogenously distributed along the
simulation cell. From the width of the velocity distribution it is
possible to define an ``\emph{effective temperature}". According
to the equilibrium probability factor
\begin{equation}
\nonumber P\propto \exp{\bigg[-\bigg(
\frac{mv_x^2}{2k_BT_{eff}^x}+\frac{mv_y^2}{2k_BT_{eff}^y}\bigg)
\bigg]},
\end{equation}

\noindent where $T_{eff}^x$ and $T_{eff}^y$ take into account also
the contributions due to the potential energy, the fluctuations of
the velocity components are related to the effective temperature
by:

\begin{eqnarray}
\nonumber &T^x_{eff}= {m<{(v'_x-<v'_x>)}^2>}/{k_B},\\
\nonumber &T^y_{eff}={m<{(v'_y-<v'_y>)}^2>}/{k_B}.
\end{eqnarray}

\noindent In our dimensionless units and our specific case this
yields:

\begin{eqnarray}
\nonumber &T'^x_{eff}=2<{(v'_x-<v'_x>)}^2>,\\
\nonumber &T'^y_{eff}=2<{v'_y}^2>,
\end{eqnarray}
\noindent respectively. The calculated effective temperatures are
reported in Table I.

\begin{widetext}

\begin{figure}
\begin{center}
\includegraphics[width=16cm]{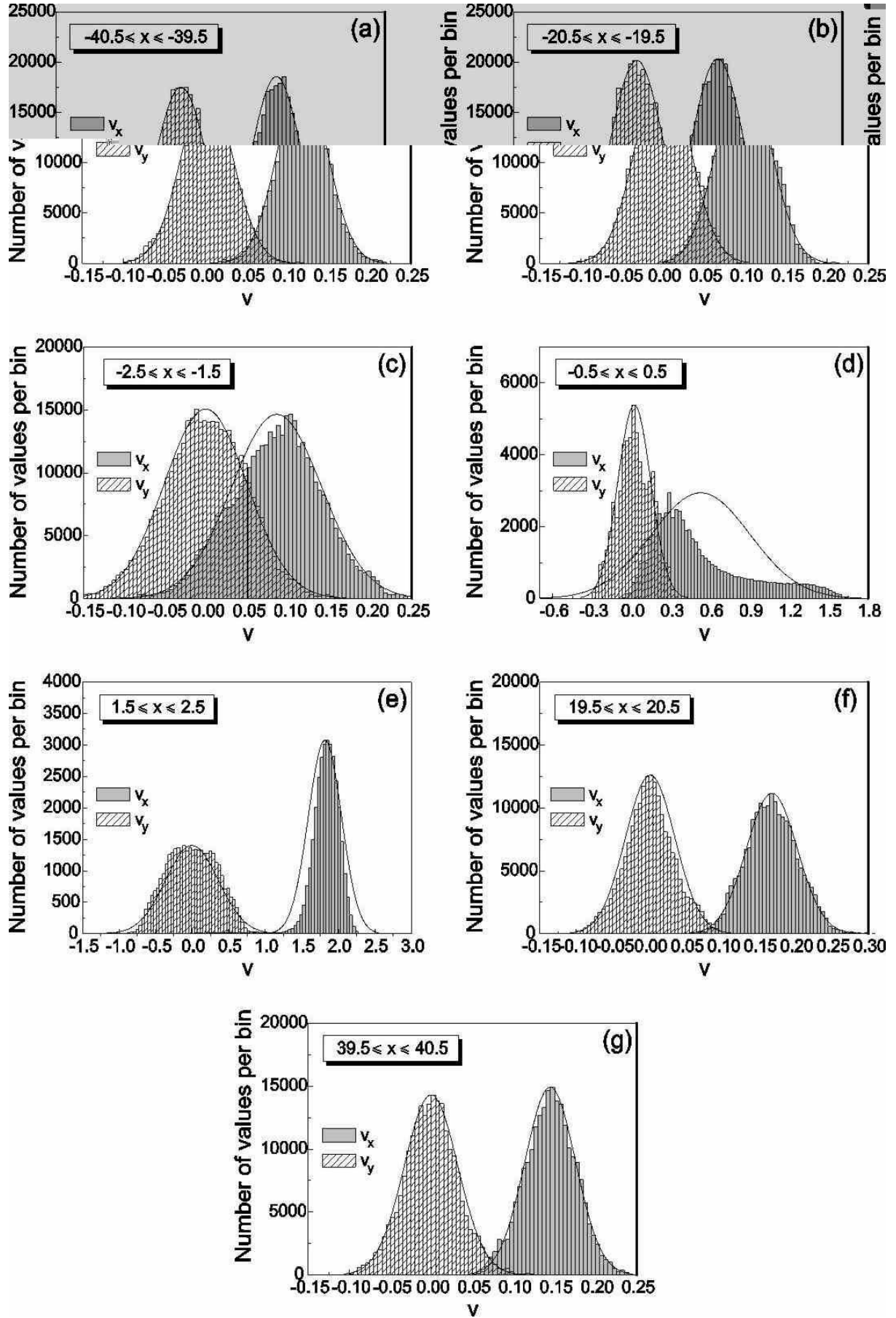}
\caption{The $v_x$ and $v_y$ distributions along the simulation
cell length for a system of $N=400$ particles, for $\kappa=1$,
$V'_0=5$ and $\alpha'=1$ and for a temperature $T'=0.002$ and
driving force $f=0.05$. The superimposed curves are the normal
distribution curves generated using the mean and standard
deviation of the data.}
\end{center}
\end{figure}

\end{widetext}

It is worth to notice that the effective temperatures are the same
as the simulation temperature in the regions far away from the
constriction barrier where the velocity fluctuations are nicely
described by a normal distribution (see Fig. 7). The effective
temperatures $T'^x_{eff}$ and $T'^y_{eff}$ increase when
approaching the constriction. In the strips $-0.5\leq x \leq 0.5$
and $1.5\leq x \leq 2.5$ both $T'^x_{eff}$ and $T'^y_{eff}\gg T'$.
This is evidently a result of the strong interaction with the
barrier which increases significantly the fluctuations in the
velocity. Actually, the spreading of the distribution of the
velocities is one to two orders of magnitude larger in the
constriction region with respect to the regions where the
constriction potential is almost zero.

\begin{table}
\begin{tabular}{|c|c|c|}
\hline  &   \bf {$T'^x_{eff}$}  &  \bf{$T'^y_{eff}$}       \\
\hline
\hline $-40.5\leq x \leq -39.5$    &  $0.0018$  & $0.0020$ \\
\hline $-20.5\leq x \leq -19.5$    &  $0.0020$     & $0.0021$ \\
\hline  $-2.5\leq x \leq -1.5$   &  $0.0060$     & $0.0051$ \\
\hline $-0.5\leq x \leq 0.5$      &  $0.28$  & $0.031$ \\
\hline $1.5\leq x \leq 2.5$     &  $0.14$     & $0.26$ \\
\hline $19.5\leq x \leq 20.5$   &  $0.0024$     & $0.0023$ \\
\hline $39.5\leq x \leq 40.5$      &  $0.0019$  & $0.0021$ \\
\hline
\end{tabular}
\caption{Effective electron temperatures corresponding to the
situation of Fig. 6(h) in the different regions studied in Fig.
7.}
\end{table}

In the elastic regime the velocity fluctuations could be fitted to
a Gaussian distribution both for $v_x$ and $v_y$, even in the
vicinity of the constriction. Around the barrier, the strong
interaction effect is felt as an increase of the effective
temperature, which is much less pronounced than in the
quasi-elastic regime.

Regarding to the noise observed in the trajectory plots, it is
essentially due to the fact that the particles merge from the
constriction with a relatively large $y$ component of the
velocity. As is seen form Fig. 7(d), the distribution of the $v_y$
spreads over a quite large range of values while passing the
constriction. Indeed, the standard deviation of the $v_y$
distribution around $x=0$ is one order of magnitude larger than in
the other regions, as already mentioned. This feature is a
consequence of the fact that very close to the barrier the
constriction force is very small (it is zero at $x=0$), the
particles proceed slowly and, thereby, they strongly undergo the
effect of the confining potential which accelerates them in the
$y$-direction, producing a significant $y$ component of the
velocity. This is also confirmed by the fact that passing the
constriction barrier a narrowing of the chain structures is always
present (see Fig. 6). Notice from Fig. 7(d,e) that the velocity
fluctuations are no longer described by a normal thermodynamic
equilibrium distribution and that in particular for $v_x$ there
are large deviations.

\begin{figure}
\begin{center}
\includegraphics[width=8cm]{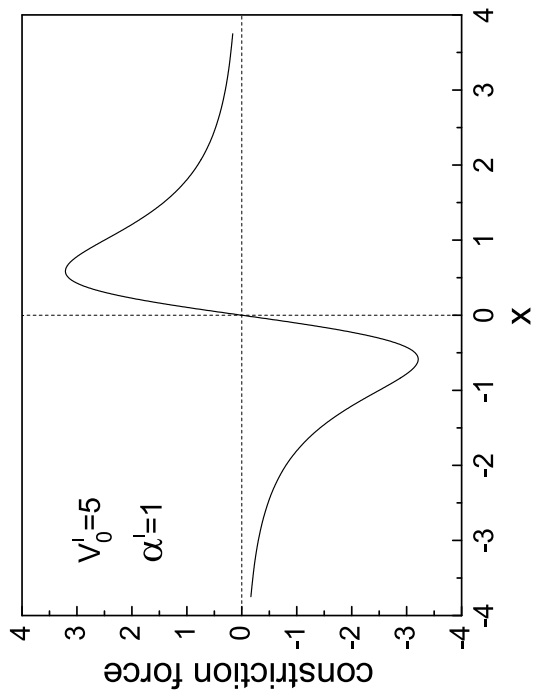}
\caption{The constriction force profile as a function of $x$ for
$V'_0=5$ and $\alpha'=1$}
\end{center}
\end{figure}

As first predicted by Fisher the elastic depinning exhibits
criticality \cite{fisher} and the velocity vs force curves scale
as $v = (f-f_c)^\beta$. This scaling has been extensively studied
in 2D CDW systems where $\beta=2/3$ \cite{narayan,myers}. It is,
however, still an open issue whether this exponent is the
signature of an universality class and whether it depends on the
particle-particle interaction. Actually, in other investigations
on elastic depinning of driven colloidal lattices the findings
were $\beta\sim0.5$ \cite{pert} and $\beta=0.92\pm0.01$
\cite{chen}. Other studies on plastic depinning with filamentary
or river-like flow have shown a velocity-force curve scaling with
$\beta=2.2$ \cite{pert}, $\beta=1.94\pm0.03$ \cite{reichhardt} for
colloids, $\beta=2.0$ for electron flow simulations in metallic
dots \cite{middleton} and $\beta=2.22$ for vortex flow in
superconductors \cite{cao1}.

As pointed out by Le Doussal and Giamarchi, for an infinite size
2D system true elastic depinning is not expected since
dislocations and defects, acting as pinning centers, should appear
at large scales \cite{giamarchi}. Both the simulations and the
experiments are, however, always for finite size systems and
consequently elastic depinning is possible where the distance
between dislocations may be larger than the system size.

In Fig. 9, we report the $v-f$ curve in the case of elastic and
quasi-elastic depinning for different number of particles, i.e.
for different chain arrangements. It should be noticed that the
critical exponent does not depend on the number of chains in the
system. For all the investigated chain configurations we obtained
on average that $\beta\simeq0.66$ in the case of homogeneous
channel flow, that is elastic depinning, and $\beta\simeq0.95$ in
the case of inhomogeneous channel flow, that is quasi-elastic
depinning. For negative value of $V'_0$ we found similar values
for the critical exponent in the elastic and quasi-elastic regime.
The value of the critical exponent could, therefore, be considered
as a clear signature of the kind of depinning. Our results are
consistent with the findings on CDW systems and colloids,
mentioned before. With increasing temperature but below the
melting temperature, we observed a broadening of the conducting
channel or some changes in the structure with some chains
collapsing (we will provide more details about this point in the
next subsection), but no significant dependence of the critical
exponent on temperature was found (within our fitting errors).

\begin{widetext}

\begin{figure}
\begin{center}
\includegraphics[width=14cm]{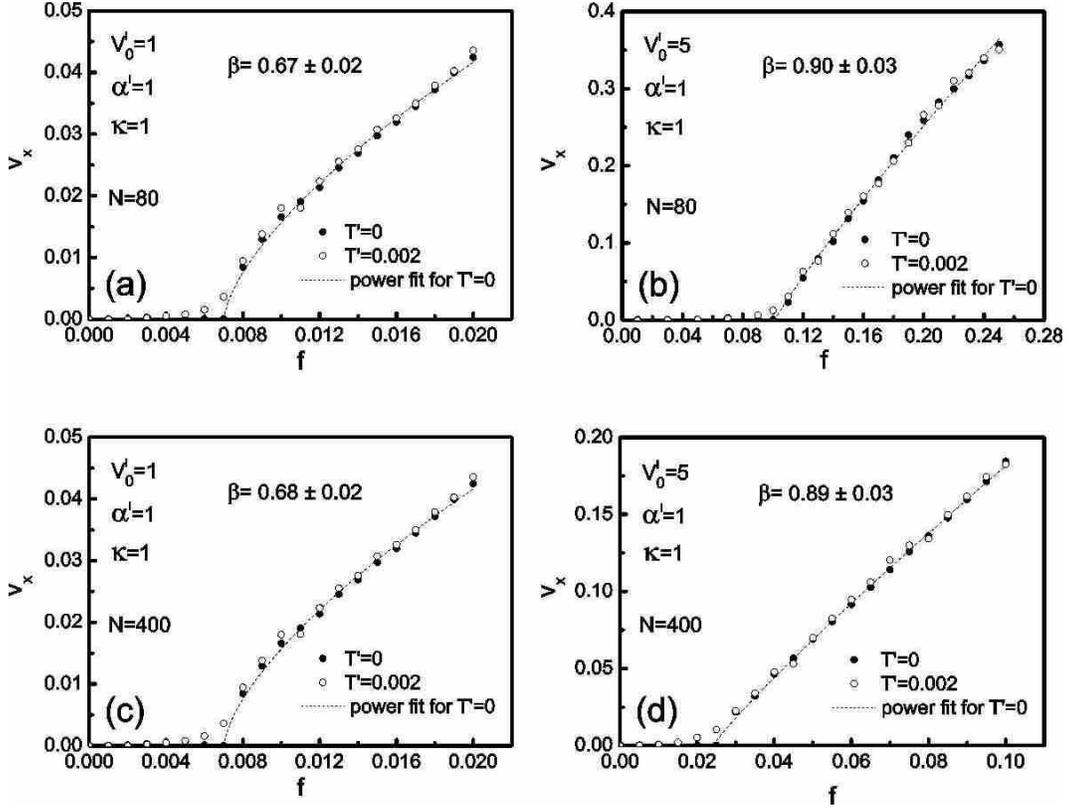}
\caption{Velocity $v_x$ vs applied drive $f$ for the elastic
depinning ((a) and (b)), for the quasi-elastic depinning ((c) and
(d)) for $\kappa=1$ and $\alpha'=1$. The dashed curves are the
best fitted power law behavior for $T'=0$. At zero temperature the
curves exhibit a sudden jump at the depinning threshold, while at
finite temperature they are smoother.  The critical exponents are
independent of the density, i.e. the number of chains.}
\end{center}
\end{figure}

\end{widetext}

The question of whether in confined systems there is an universal
exponent for elastic and quasi-elastic depinning cannot be
answered conclusively. We found that the critical exponent are not
affected by the value of $\kappa$, as it can be seen in Table II:
going from $\kappa=0.2$ (nearly Coulomb interaction) up to
$\kappa=5$ (short range interaction), the value for the critical
exponent stays the same. Although we found that the critical
behavior is independent of the particle-particle interaction,
further investigations with other kind of pinning potentials are
required in order to affirm that the elastic, quasi-elastic and
plastic depinning belong to different universality classes.

In our simulations neither in case of elastic nor quasi-elastic
depinning history dependence was found. The velocity vs applied
drive are not hysteretic, and we obtained the same result for
increasing and decreasing values of $f$.

Finally, in the case of very small $\alpha$, that is for a wide
constriction interaction, because of the density gradient
producing the coexistence of different chain structures, we always
observed quasi-elastic depinning.

\begin{table}
\begin{tabular}{|c|c|c|c|}
\hline $\kappa$ &  $V'_0$  & $\beta$       & Depinning\\
\hline
\hline $0.2$    &  $0.25$  & $0.68\pm0.04$ & elastic \\
\hline $0.2$    &  $1$     & $0.70\pm0.06$ & elastic \\
\hline $0.2$    &  $5$     & $0.95\pm0.04$ & quasi-elastic \\
\hline $1$      &  $0.25$  & $0.64\pm0.03$ & elastic \\
\hline $1$      &  $1$     & $0.67\pm0.02$ & elastic \\
\hline $1$      &  $5$     & $0.92\pm0.05$ & quasi-elastic \\
\hline $1.5$    &  $0.25$  & $0.63\pm0.05$ & elastic \\
\hline $1.5$    &  $1$     & $0.68\pm0.04$ & elastic \\
\hline $1.5$    &  $5$     & $0.95\pm0.06$ & quasi-elastic \\
\hline $2$      &  $0.25$  & $0.65\pm0.04$ & elastic \\
\hline $2$      &  $1$     & $0.66\pm0.03$ & elastic \\
\hline $2$      &  $5$     & $0.96\pm0.03$ & quasi-elastic \\
\hline $3$      &  $0.25$  & $0.65\pm0.03$ & elastic \\
\hline $3$      &  $1$     & $0.95\pm0.06$ & quasi-elastic \\
\hline $3$      &  $5$     & $0.97\pm0.05$ & quasi-elastic \\
\hline $4$      &  $0.25$  & $0.68\pm0.04$ & elastic \\
\hline $4$      &  $1$     & $0.94\pm0.04$ & quasi-elastic \\
\hline $4$      &  $5$     & $0.91\pm0.05$ & quasi-elastic \\
\hline $5$      &  $0.25$  & $0.67\pm0.04$ & elastic \\
\hline $5$      &  $1$     & $0.98\pm0.06$ & quasi-elastic \\
\hline $5$      &  $5$     & $1.02\pm0.08$ & quasi-elastic \\
\hline
\end{tabular}
\caption{The critical exponent $\beta$ for different values of the
inverse screening length $\kappa$ and the constriction barrier
height $V'_0$.}
\end{table}

\subsection{Crossover from elastic to quasi-elastic depinning}

It is interesting to investigate the values of $\beta$, or in
other word the kind of flow, as a function of $V'_0$. For
colloidal systems a sharp crossover from elastic to plastic
depinning was found with increasing strength of the substrate
disorder, accompanied by a sharp increase in the depinning
critical force \cite{reichhardt}. Carpentier and Le Doussal
studied theoretically the effect of quenched disorder on the order
and melting of 2D lattices and found a sharp crossover from the
ordered Bragg glass (where there are no defects) to a disordered
state \cite{carpentier}. They also predicted that the depinning
threshold increases at the order to disorder transition due to the
softening of the lattice, which allows the particles to better
adjust to the substrate. A similar mechanism could account for the
peak effect observed in vortex matter in superconductors
\cite{bhatta}, in which the depinning threshold rises abruptly
with increasing applied magnetic field.

We found a crossover from elastic to quasi-elastic depinning as
the barrier height of the constriction is increased. This is
analogous to the crossover from the elastic to plastic flow
encountered in other systems. As can be seen in Fig. 10, the
behavior of $\beta$ as a function of $V'_0$ is almost step-like,
and the crossover takes place in a narrow range of $V'_0$ values.
We also observed increasing values of the critical threshold
$f_c$. It is beyond the scope of this paper to determine whether
the elastic to quasi-elastic crossover is a first or second order
transition and how temperature influences this transition.
However, the relative smoothness of the curves in Fig. 10 suggests
a possible second order transition. Furthermore, the effect of
increasing temperature should reasonably result in a shift to
lower values of $f_c$ or $V'_0$ for the transition from elastic to
quasi-elastic flow.

\begin{figure}
\begin{center}
\includegraphics[width=7.5cm]{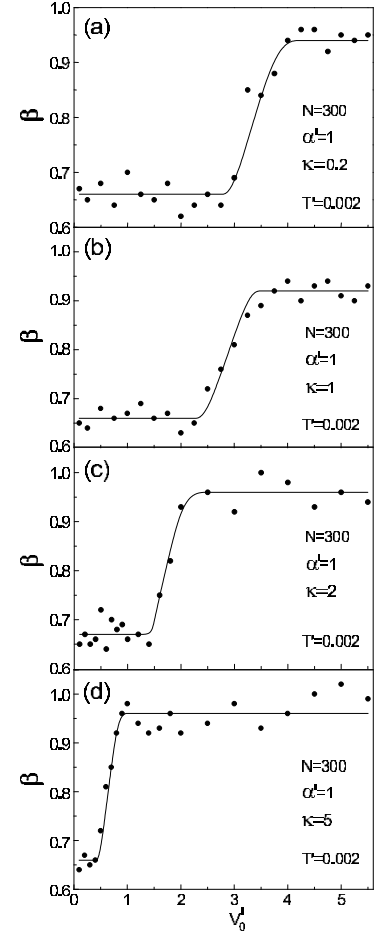}
\caption{The critical exponent $\beta$ as a function of the
constriction barrier height $V'_0$ for different values of the
inverse screening length: (a)$\kappa=0.2$, (b) $\kappa=1$ and (c)
$\kappa=5$. With increasing values of $\kappa$ the crossover
region shifts to lower values of $V'_0$. The solid lines are
guides to the eye.}
\end{center}
\end{figure}

It is evident from Fig. 10 that the crossover shifts towards lower
values of the potential barrier as the inverse screening length is
increased, which means that the particles with stronger
interparticle interactions can flow in a more ordered way.

The values of the critical exponent $\beta$ shown in Fig. 10 can
be fitted by the curve $a\tanh{(b(V'_0-\overline{V}'_0))+c}$,
where $a$, $b$, $c$ and $\overline{V}'_0$ are the fitting
parameters. In particular, the sum $a+c$ gives the value of
$\beta$ for the case of quasi-elastic depinning, while the
difference $c-a$ gives the value of $\beta$ for the case of
elastic depinnig, $\overline{V}'_0$ can be identified with the
value of the constriction height for which the crossover from
elastic to quasi-elastic depinning takes place and $b$ is related
to the sharpness of the transition or, to be more precise, it is
related to the inverse width of the transition region. The results
of the fits are reported in Table III.

\begin{table}
\begin{tabular}{|c|c|c|c|c|}
\hline $\kappa$ &  $a$  & $b$ & $c$ & $\overline{V}'_0$\\
\hline
\hline $0.2$    &  $0.144\pm0.007$  & $2.1\pm0.6$ & $0.799\pm0.007$ & $3.28\pm0.06$\\
\hline $1$      &  $0.131\pm0.005$  & $2.4\pm0.4$ & $0.790\pm0.004$ & $2.90\pm0.05$\\
\hline $1.5$    &  $0.139\pm0.005$  & $3.2\pm0.4$ & $0.796\pm0.005$ & $2.12\pm0.06$\\
\hline $2$      &  $0.143\pm0.006$  & $4.2\pm1.2$ & $0.813\pm0.006$ & $1.78\pm0.04$\\
\hline $3$      &  $0.152\pm0.005$  & $4.8\pm1.3$ & $0.811\pm0.007$ & $1.24\pm0.05$\\
\hline $4$      &  $0.146\pm0.007$  & $5.2\pm1.3$ & $0.802\pm0.008$ & $0.94\pm0.06$\\
\hline $5$      &  $0.160\pm0.008$  & $5.4\pm1.4$ &  $0.80\pm0.01$  & $0.61\pm0.03$\\
\hline
\end{tabular}
\caption{The fitting parameters for the crossover curves for
different values of the inverse screening length $\kappa$.}
\end{table}

As it can be seen from Table III, the constriction height for
which the quasi-elastic regime is established is a decreasing
function of the inverse screening length $\kappa$, while the
sharpness of the transition is an increasing function of $\kappa$
(see also Fig. 11). Again it confirms that for long range
interactions the particles can flow more orderly. The errors in
the fitting parameters are small for $a$, $c$ and
$\overline{V}'_0$ and relatively large for $b$.

In Fig. 11, the values of $\overline{V}'_0$ and $b$ as a function
of $\kappa$ are reported. The curve
$\overline{V}'_0\equiv\overline{V}'_0(\kappa)$ is well fitted by a
Lorentzian $1/(p+q\kappa^2)$ with $p=0.302\pm0.005$ and
$q=0.058\pm0.002$ and the inverse width (see inset of Fig. 11) by
the Pad\'{e} approximation $(g+h\kappa^2)/(1+l\kappa^2)$ with
$g=1.9\pm0.2$, $h=1.4\pm0.4$ and $l=0.23\pm0.07$.

\begin{figure}
\begin{center}
\includegraphics[width=7.5cm]{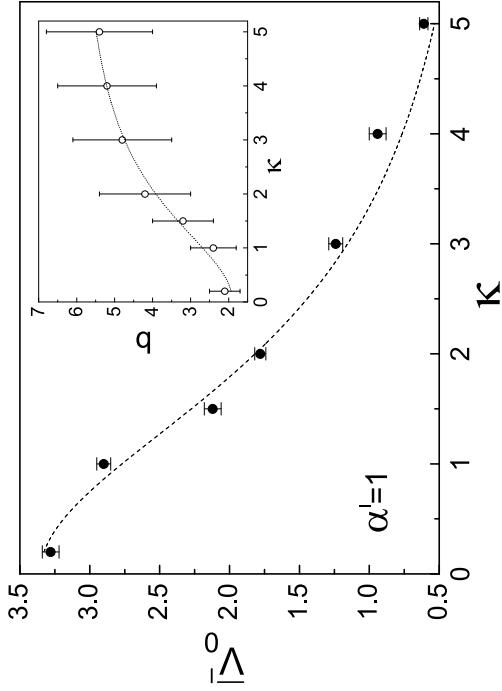}
\caption{The transition points $\overline{V}'_0$ as a function of
the inverse screening length $\kappa$. The dashed line is the
fitting curve $1/(p+q\kappa^2)$. The inset shows the behavior of
the inverse width of the transition region as a function of
$\kappa$. The dotted line is the fitting curve
$(g+h\kappa^2)/(1+l\kappa^2)$.}
\end{center}
\end{figure}

We should stress that the physics behind the crossover from
elastic to quasi-elastic depinning is different from the case of
quenched disorder, where with increasing disorder strength the
ordered structure is softened and particles can better adjust to
the substrate. In our case the accumulation of particles in the
vicinity of the constriction barrier and their mutual repulsion
give rise for high values of $V'_0$ to a complex arrangement of
conducting channels, in which the nearest neighbors of each
particle change, i.e to the impossibility of elastic flow.

\subsection{Conductivity}
According to Ohm's law the current density of a classical system
of charged particles is proportional to the applied electric
field:

\begin{equation}
\vec{j}=\vec{\sigma}\cdot\vec{E},
\end{equation}

\noindent where $\vec{\sigma}$ is the specific conductivity. The
conductivity can be in general expressed as a second rank tensor.
Because of the geometry of the investigated system and because the
driving is in the $x$-direction, we are interested only in
$\sigma_{xx}$, which we will refer in what follows as simply
$\sigma$.

From the definition of $j$ and from Eq. (6a) it follows, in
dimensionless units:

\begin{equation}
\sigma'=\widetilde{n}_{e}\frac{v'_x}{f}=\frac{\widetilde{n}_{e}}{\gamma'},
\end{equation}

\noindent where $\sigma'=\sigma /(q^2/m \omega_0 r_0)$. As the
definition of total density is not always an accurate one for our
system, as mentioned above, we investigated the ratio $v'_x/f$,
which is directly proportional to $\sigma'$ through
$\widetilde{n}_{e}$ and which is a constant equal to $1/\gamma'$,
according to Eq. (8). In Fig. 12 the results of our calculations
are reported for different values of the friction.

\begin{figure}
\begin{center}
\includegraphics[width=8cm]{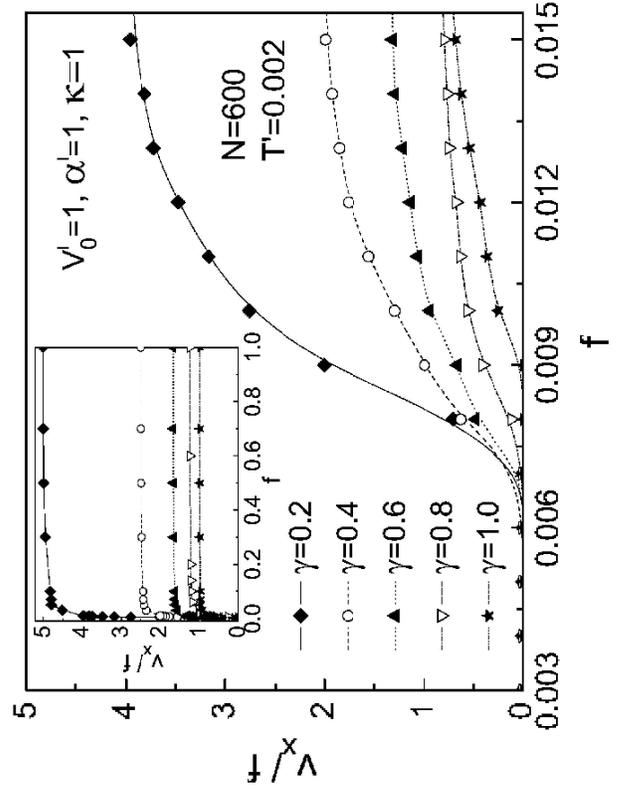}
\caption{The ratio between velocity and driving force as a
function of the drive strength for different values of the
friction coefficient. The conductivity is non-Ohmic in a narrow
region above the depinning threshold. The inset shows that for
larger values of $f$ Ohm's law is fulfilled with the conductivity
as a constant. The curves are guides to the eye.}
\end{center}
\end{figure}

When the particles are pinned the conductivity is obviously zero.
It is interesting to notice that after the depinning threshold,
there is a narrow region where the conductivity shows non-Ohmic
features, going from zero to the saturation value $1/\gamma'$.
Afterwards when the drive is the leading effect in the equations
of motion, the particles behave as a classical Ohmic conductor.
This behavior is independent of the number of particles and of the
height of the constriction barrier. For higher values of $V'_0$
the non-Ohmic conductivity region is enlarged.

Studying the conductivity as a function of temperature some
interesting features were observed. We investigated a rather wide
range of temperatures from 0.001 to 0.018. From Ref.
\cite{piacente} we know that for the single chain configuration in
the absence of driving and constriction potential, the melting
temperature is arbitrarily low, while in the multiF-chain
configuration it is $T'_m\sim0.015$. The results of our
calculations, for a driving force $f=0.05$ and for different
values of $V'_0$, are sketched in Fig. 13.

\begin{figure}
\begin{center}
\includegraphics[width=8cm]{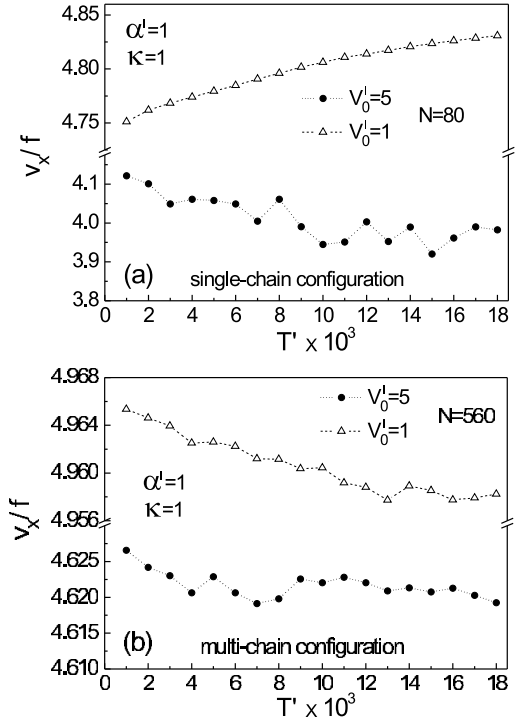}
\caption{Dependence of the conductivity on the temperature, for
(a) a single-chain structure and (b) a multi-chain structure. For
weak values of the potential barrier the conductivity of the
single-chain structure is an increasing function of temperature,
in all the other cases it shows a decreasing trend.}
\end{center}
\end{figure}

The behavior in the single chain configuration shows substantial
differences from the multi-chain one. First of all, for weak
values of $V'_0$, the conductivity is an increasing function of
$T'$ in the single chain case, while it is decreasing in the
multi-chain case. This means that when the number of particles is
small, that is the electrostatic interaction is not too strong,
the thermal motion aids the particles to overcome the potential
barrier and does not act as a disturbance, while in the case of a
large number of particles, the possibility of overcoming the
barrier is sustained by the electrostatic repulsion and the
thermal agitation is a dissipative factor. This is an ulterior
confirmation that the dynamics of the system under investigation
is a very complex interplay of driving, electrostatic interaction,
repulsion from the constriction, thermal fluctuations and
confinement.

In general, in a classical model of conduction the conductivity is
expected to be a decreasing function of temperature. For large
values of the constriction barrier height either in the case of
single chain and multi-chain structures, we found that the
conductivity is not a monotonic function of temperature, although
it shows a decreasing trend. The presence of structure in the
$v_x/f$ curve vs $T'$ can be explained by the fact that with
increasing temperature some of the channels formed for high $V'_0$
can collapse together to form new channels, as mentioned in
previous subsection. This is confirmed by the analysis of the
configurational energy per particle as a function of temperature,
which is plotted in Fig. 14.

\begin{figure}
\begin{center}
\includegraphics[width=8cm]{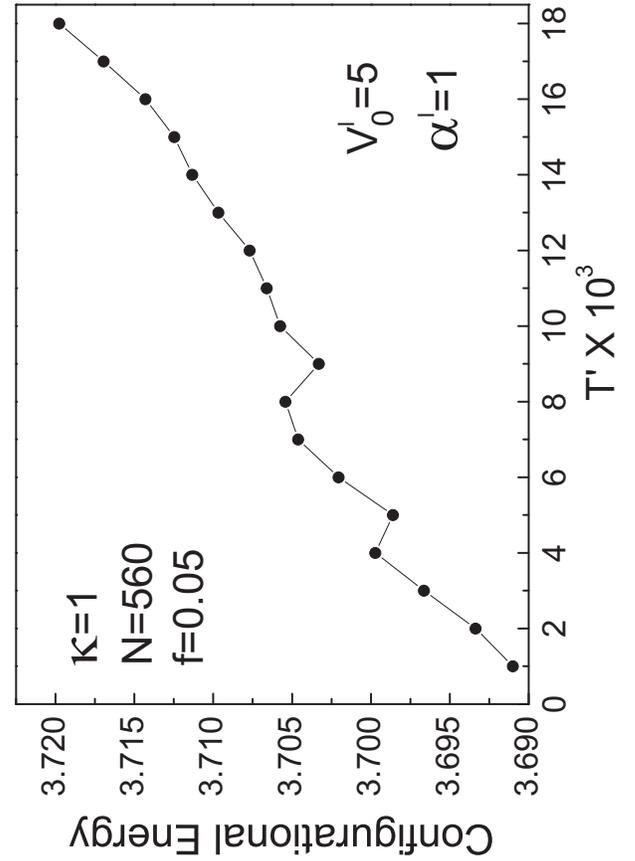}
\caption{The average configurational energy per particle as a
function of the temperature for a multi-chain configuration.}
\end{center}
\end{figure}

The average potential energy exhibits the same temperature
dependance as the conductivity, while in the case of a weak
constriction barrier height the configurational energy per
particle increases linearly with temperature. Another factor,
which is also responsible for that non monotonicity, is the fact
that, in the case of high barrier values, a density gradient is
present and the melting is not homogenous, thus some parts of the
system can be in the liquid state while others are still in the
solid state, giving rise to complex phenomena in the transport
properties.

\section{Other Dynamical properties}

For high values of the driving force the system shows the
phenomenon of \emph{dynamical reordering}. When the driving force
is large the system, even in case of a high constriction barrier,
can flow in an ordered channel structure. This is shown in Fig.
15. It is interesting to compare the trajectories of Fig. 15 with
the one of Fig. 6(j). Above the depinning threshold (Fig. 6(j))
the channel structures are not homogenous, with increasing the
drive (Fig. 15) an ordered moving structure is reached again.

\begin{figure}
\begin{center}
\includegraphics[width=9.5cm]{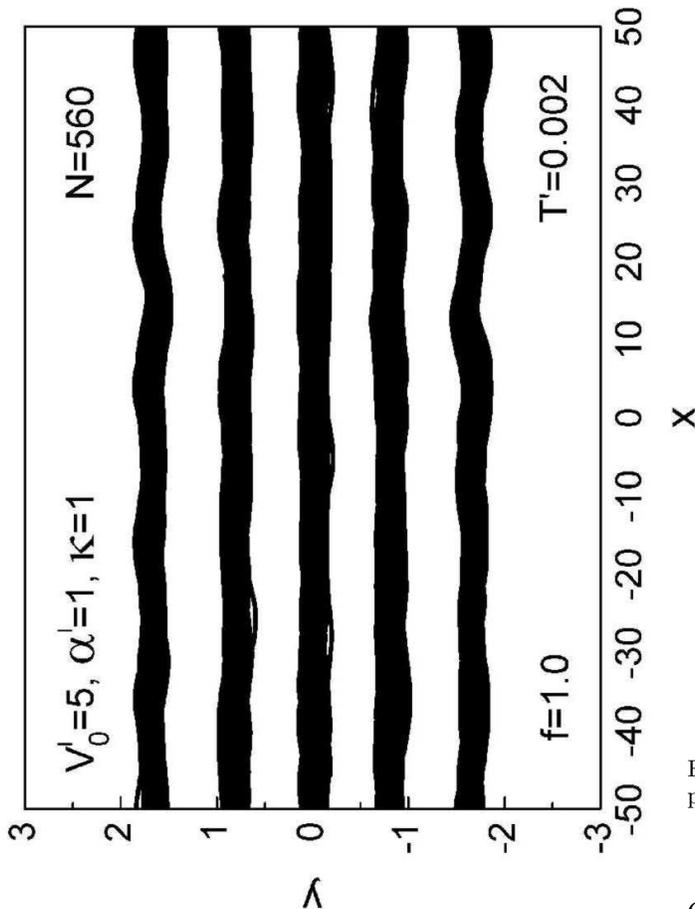}
\caption{Trajectories of a system of N=560 particles for a high
value of the driving force. The five channel structure is attained
over the whole wire (cfr. Fig. 7(j)).}
\end{center}
\end{figure}

This is a well known phenomenon. Indeed, it was observed
experimentally for vortex lattices in type II superconductors
\cite{yaron} and for colloids \cite{helle}. The interplay between
dynamical reordering and melting in mesoscopic channels was
recently studied experimentally in the case of vortices, providing
the first conclusive evidence for a velocity dependent melting
transition \cite{kes}. The dynamical reordering was also
investigated theoretically for CDW systems \cite{fisher}. The
dynamical reordering for such systems originates from the fact
that the applied driving force tilts the pinning potential thereby
reducing the pinning strength. When a large enough force is
applied, the particles depin and then flow quite orderly. The same
mechanism is responsible for the dynamical reordering of the
studied system, with the difference that the tilted potential is
in this case the constriciton potential.

It is worth to study the ratio between kinetic and potential
energy, averaged at every simulation step, as a function of the
temperature. From Fig. 16 it is evident that the kinetic energy
increases faster than the configurational energy with temperature,
which is an expected result. What is interesting is the fact that
the fitting curve is of the type $y=c/(1+dx)$. The fit is
excellent with very small errors in the fitting parameters $c$ and
$d$.

\begin{figure}
\begin{center}
\includegraphics[width=8cm]{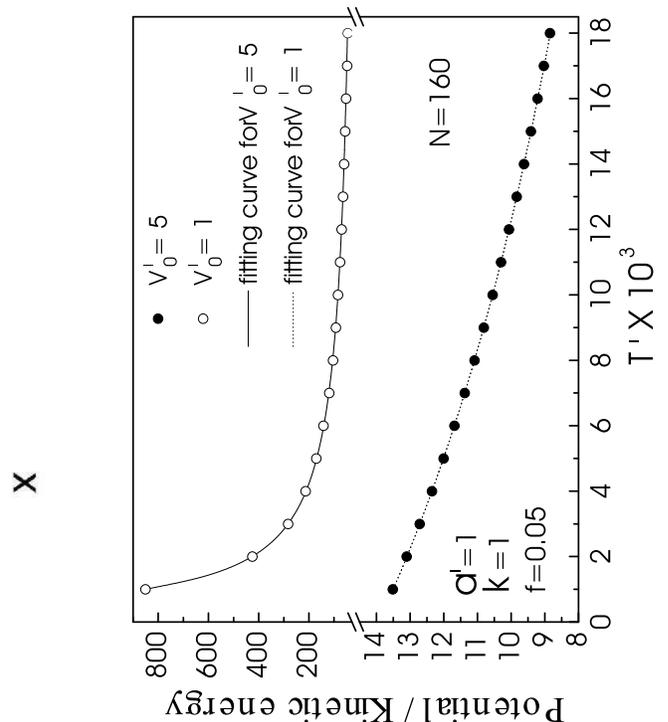}
\caption{The ratio between potential and kinetic energy per
particle as a function of temperature.}
\end{center}
\end{figure}

Finally, we also investigated the distribution of the velocity
$v_x$ as a function of the distance from the $x$ axis, or in other
words we studied the velocity for each conducting channel when the
system flows orderly. Motivated by the experimental findings of
Ref. \cite{glasson}, that for a chain system of electrons on
liquid helium, the particles in the external chains have higher
velocity than the particles in the internal chains, we tried to
clarify whether in our system a similar behavior occurs. Our
findings are in contrast with the one of Ref. \cite{glasson},
actually we found that the internal chains have on average a
velocity which is 5\% higher than the external chains. However,
this discrepancy can be explained by the circumstance that the
pinning mechanism are different for the two systems: the coupling
between electrons and ripplons in case of Ref. \cite{glasson} and
the constriction potential in our case, and by the fact that the
confinement potentials are not exactly the same in the two
systems.

\section{Comparison with other driven systems}

A rather general theory of periodic structures in a random pinning
potential under the action of an external driving force was
developed by Le Doussal and Giamarchi \cite{giamarchi}. Their
findings were that the periodicity in the direction transverse to
the motion leads to a different class of driven system: the moving
glasses, with the decay of the translational long-range order as a
power law. Similar considerations can be made for our system as
well, but with the important difference that because of the
confining potential and the constriction potential the periodicity
is broken both in the $x$ and $y$-direction. For weak constriction
barrier height long-range translational order is present in the
$x$-direction, but it is softened when the system is moving and
the temperature is increased. For infinite systems one of the
consequences of periodicity in the transverse direction to the
motion is that particles flow along static channels for
uncorrelated and weak disorder and that there are barriers to
transverse motion. In our confined system the barriers to
transverse motion are an effect resulting from confinement instead
of periodicity. Most of the studies on driven lattices or glasses
show that, at finite but low temperature, the channels broaden and
strong non-linear effects exist in the response to the applied
drive, though the asymptotic behavior is found to be linear, which
is, indeed, what we found as well.

In infinite moving systems with random pinning centers, depending
on the strength of that disorder, two kinds of flow are possible:
(i) the \emph{elastic} one, where all the particles move keeping
their neighbors, and (ii) the \emph{plastic} one, where part of
the particles are moving in river-like or filamentary structures
and part is pinned. There is a sharp crossover from the elastic to
the plastic flow, related to an order-disorder transition. In our
system we found that two kinds of flow are possible: (i) the
\emph{elastic} flow, where all the particles move orderly and the
nearest neighbors are preserved, and (ii) the \emph{quasi-elastic}
flow, where all the particles move together, but creating a
complex net of conducting channels, for which the neighbors are
not kept. We also found a continuous crossover from the elastic to
quasi-elastic flow. It is important to stress that this difference
is closely related to the different pinning potentials considered.
To be more precise, in the case we investigated the pinning is due
to a constraint rather than an actual pinning potential. That is
the reason why we did not observe plastic flow, because particles
cannot be strongly attracted and pinned by any pinning center.

It is remarkable that for our system we also found that in the
case of elastic depinning the velocity vs driving force curve
scales as $v\propto(f-f_c)^\beta$, with $\beta\sim0.66$, which is
in agreement with most of the theoretical and numerical works for
infinite systems exhibiting elastic flow, where $\beta=2/3$. Thus,
the value of the critical exponent $\beta=2/3$ seems actually the
signature of elastic depinning independently of the presence of
confinement and the type of pinning potential. Naturally, in order
to affirm this definitely more investigations are required with
different topologies and potentials. Furthermore, in the case of
quasi-elastic depinning we found a critical exponent
$\beta\sim0.95$, which is an intermediate value between the case
of elastic and plastic flow, where the experimental findings give
$\beta\sim2$. This leads us to the conclusion that the
quasi-elastic depinning is an intermediate regime between elastic
and plastic depinning.

Finally, for the elastic regime the previous theoretical
investigations followed essential two approaches: (i) elastic
theory with renormalization group techniques
\cite{giamarchi,carpentier} and (ii) perturbation theory in $1/v$
\cite{sneddon,larkin,schmidt}. The first one explains the flowing
channel structures and their mutual interactions, while the second
one elucidates the $v-f$ characteristics and the criticality in
the depinning. Despite the number of experimental and numerical
data a detailed theoretical understanding of plastic motion
remains still a challenge \cite{watson}.

\section{Conclusions}

We studied the ground state and the dynamical properties of a
classical Q1D infinite system of particles interacting through a
Yukawa-type potential and with a Lorentzian shaped constriction
potential. The system is confined in one direction by a parabolic
potential. By MC simulations we found that at $T=0$ the particles
arrange themselves in a chain-like system, where the number of
chains are a function of the number of particles, i.e. the
density. Depending on the height and on the interaction range of
the constriction barrier, a density gradient in the chain
configuration is present near the constriction.

We studied the response of the system when an external driving
force is applied in the not confined direction. We performed
Langevin molecular dynamics simulations with periodic boundary
conditions in the not confined direction and open conditions in
the confined direction for different values of the driving force
and for different temperatures. We found that the constriction
barrier and the friction pins the particles up to a critical value
of the driving force. The pinned phase is a new static phase, with
particles accumulating in the neighborhood of the constriction and
arranging themselves in such a way to balance the external drive.
For values of the driving force which are higher than the critical
threshold, the particles can overcome the potential barrier and
the system depins. We analyzed in detail the depinning phenomenon
and we found that the system can depin elastically or
quasi-elastically depending on the strength of the constriction
potential. The quasi-elastic flow is a new regime, where particles
move together without keeping their neighbors.

In the case of elastic flow the chain-like structure, formed at
$T=0$ in the absence of external drive, is preserved, while in the
case of quasi-elastic flow it is destroyed and a complex net of
conducting channels is created. The elastic depinning is
characterized by a critical exponent, which is on average
$\beta\sim0.66$ and does not depend on the number of chains. This
is in excellent agreement with the theoretical and numerical
findings on 2D systems exhibiting elastic depinning. The
quasi-elastic depinning state has a critical exponent
$\beta\sim0.95$. We demonstrated that the values of the critical
exponent are independent of the range (i.e. screening length) of
the interparticle interaction. But the crossover between elastic
and quasi-elastic flow depends on the kind of interparticle
interaction.

Furthermore, we showed that the dc conductivity is zero in the
pinned regime, it has non-Ohmic characteristics after the
activation of the motion and then it is constant, in other words
the system has a non linear response to the applied drive. The
linear regime is attained as the asymptotic behavior. The
dependence of the conductivity with temperature and strength of
the constriction was also investigated. We found that in the
single chain configuration for low height of the constriction, the
conductivity is an increasing function of temperature, while in
the multi-chain configuration it is a decreasing function, as
expected. For high constriction barrier height, the conductivity
has no longer a monotonic behavior, although it has a decreasing
trend. In these cases some structures are present in the
conductivity vs temperature curve, signaling the circumstance that
some channels collapse or some parts of the system have already
undergone the transition form the solid to the liquid state.
Finally, for large values of the external driving force even in
the case of high constriction barrier, the particles can flow
orderly in a well defined channel structure, because the drive
tilts the contriction  potential, thus reducing the pinning
strength, that is the system exhibits the phenomenon of dynamical
reordering.

\section{Acknowledgments}
This work was supported in part by the European Community's Human
Potential Programme under contract HPRN-CT-2000-00157 "Surface
Electrons" and the Flemish Science Foundation (FWO-Vl). We thank
Dr. I. Schweigert for interesting discussions.

\end{document}